\documentclass[preprint,showpacs,showkeys,preprintnumbers,amsmath]{revtex4}
\usepackage{amsmath}
\usepackage{graphicx}
\usepackage{dcolumn}
\usepackage{bm}

\begin{document}

\title{Exactly Integrable Analogue of a One-dimensional
Gravitating System}

\author{Bruce N. Miller\email{B.Miller@tcu.edu}}
\author{Kenneth R. Yawn\email{kenneth.r.yawn@lmco.com}}
\author{Bill Maier \email{ciric@gte.net}}
\address{Department of Physics and Astronomy, Texas
Christian University,\\Fort Worth, Texas 76129}


\begin{abstract}
Exchange symmetry in acceleration partitions the configuration
space of an $N$ particle, one-dimensional, gravitational system
into $N!$ equivalent cells. We take advantage of the resulting
small angular extent of each cell to construct a related
integrable version of the system that takes the form of a central
force problem in $N-1$ dimensions. The properties of the latter,
including the construction of trajectories, as well as several
continuum limits, are developed. Dynamical simulation is employed
to compare the two models. For a class of initial conditions,
excellent agreement is observed.
\end{abstract}

\maketitle

One dimensional systems have played a central role in the
development of physical models. Because of their relative
simplicity, they form the starting point for the study of a wide
variety of interactions. One-dimensional models of gases have been
studied extensively in statistical mechanics and one-dimensional
lattice models have been used in solid state physics to study
metal alloys, magnetic spin systems, glasses, phonon propagation,
electronic bands and a host of other systems. One dimensional
gravitational systems have been used to investigate relaxation to
equilibrium, phase transitions, the gravothermal catastrophe, and
cosmology\cite{vlasovia}. An important example is the system
consisting of a one-dimensional chain of coupled, nonlinear,
oscillators. Known as the Fermi-Pasta-Ulam (FPU)
problem\cite{FPU}, it has had an enormous impact on the
development of dynamics in the last fifty years. Its failure to
come to thermal equilibrium in simulations on the earliest
electronic computers stimulated intense research in the theory of
conservative dynamical systems and resulted in major breakthroughs
such as the KAM theorem and developments in soliton theory
\cite{FPU}. The analogous system in astronomy consists of $N$
parallel planar mass sheets of infinite extent that are restricted
to move perpendicular to their surface. Like the FPU problem, this
is a nonlinear, one-dimensional system. The only forces on a given
sheet (or particle) are gravitational, and the system obeys the
Poisson equation in one dimension. Since each particle experiences
a constant acceleration until it encounters a neighboring
particle, the motion can be expressed analytically and it is
straightforward to simulate the system on the computer
\cite{heap}.

The one dimensional self-gravitating system (OGS) was the first
gravitational system to be investigated numerically \cite{CL1}
and, like the FPU system, it resists relaxing to statistical
equilibrium. Early studies showed that the system virialized
within 50 or so characteristic system crossing times, $\tau _{c}$,
\cite{CL1} as a consequence of ''violent relaxation''
\cite{lynden1}. Following this period the system appears to
approximate a stationary Vlasov state which evolves extremely
slowly, perhaps on the order of millions of $\tau _{c}$. Evidence
of the approach to thermal equilibrium has only been recently
established for a two component system \cite{yawn3,2mass}. For an
extensive review of the history of the relevant investigations see
\cite{vlasovia,2mass}. In this paper we offer a possible
explanation for the incredibly slow exploration of phase space
these systems exhibit. We demonstrate analytically and with
numerical simulation that an exactly integrable Hamiltonian system
(EIS) exists which closely represents the OGS as its population
increases. Further we derive a continuum (Vlasov) limit for the
EIS and show that it is characterized by two integral invariants,
as well as an exactly periodic radius. We also show that the
continuum model can be extended to the pair space and provide
information concerning correlations in position and velocity.

Consider an OGS consisting of $N$ parallel mass sheets (particles) of equal
mass surface density $m$ and let the real $x$ axis be perpendicular to their
surface. Label the sheets in terms of their increasing ordering from left to
right so their positions are $x_{j}$ with $x_{j+1}>x_{j}(j=1,...,N-1)$ .
Then, from Gauss' law, the acceleration $A_{j}$ experienced by particle $j$
is a constant proportional to the net difference in the number of particles
to its right and left:

\begin{equation}\label{acc}
A_{j}=2\pi mG(N+1-2j).
\end{equation}

As the system evolves, each particle undergoes uniform acceleration until a
pair of particles crosses, at which time their accelerations are exchanged.
Since the system is isolated from external forces, the total momentum and
energy are conserved, while the center of mass moves with uniform velocity.
Without loss of generality we may assume we are in a frame where the total
momentum is zero and the center of mass is at rest. To establish a
connection with the Vlasov limit, it is more convenient to represent the
system in terms of positions and velocities ($x_{j},v_{j}$) of the
particles. Thus

\begin{equation}\label{constraints}
\sum_{1}^{N}x_{j}=0,\ \sum_{1}^{N}v_{j}=0,\ E=\frac{m}{2}
\sum_{1}^{N}v_{j}^{2}+2\pi m^{2}G\sum_{i<j}\left|
x_{i}-x_{j}\right|
\end{equation}

where $E$ is the total energy. It is also convenient to introduce a system
of unit vectors , $\mathbf{e}_{j}$ , in the $N$ dimensional configuration
space. Then we can represent the system in terms of $N$ dimensional
position, velocity (in the tangent space), and acceleration vectors,

\begin{equation}\label{Ndim}
\mathbf{x}=\sum_{1}^{N}x_{j}\mathbf{e}_{j}\ ,\
\mathbf{v}=\sum_{1}^{N}v_{j} \mathbf{e}_{j}\ ,\
\mathbf{A}=\sum_{1}^{N}A_{j}\mathbf{e}_{j}.
\end{equation}

In this context, the system is represented by a single point
moving with constant acceleration until a crossing occurs, whereby
two of the acceleration components are exchanged; i.e. the actual
components of $\mathbf{A}$ depend on the ordering of the particle
positions$\mathbf{.}$ Letting
$\mathbf{e}_{c}\mathbf{=}\frac{1}{N}\sum_{j}\mathbf{e}_{j}$, the
constraints on the total momentum and the center of mass can be
expressed by $\mathbf{v\cdot e}_{c}\mathbf{=x\cdot
e}_{c}\mathbf{=0}$. As a result of the dynamical constraints
(including energy conservation), the trajectories lie on a $2N-3$
hyper-surface of the $2N$ dimensional phase space. We can also
introduce a generalized angular momentum tensor as a two form, or
dyadic in the older formalism \cite{geom}, and its square:

\begin{equation}\label{AngMom}
\mathbf{L}=\mathbf{x}\wedge
m\mathbf{v}=m\sum_{i,j=1}^{N}x_{i}v_{j}\mathbf{e}_{i}\wedge
\mathbf{e}_{j}=\frac{m}{\sqrt{2}}
\sum_{i,j=1}^{N}(x_{i}v_{j}-x_{j}v_{i})\mathbf{e}_{i}\mathbf{e}_{j}\
\end{equation}
\begin{equation}
L^{2}=\frac{m^{2}}{2}\sum_{i,j=1}^{N}(x_{i}v_{j}-x_{j}v_{i})^{2}.
\end{equation}

It is straightforward to show that, for $N=3$, these definitions
reduce to the familiar equations of three dimensional orbital
mechanics. For completeness, and to compare with our further
development below, we mention that the system has a well defined
continuum (Vlasov) limit. By projecting the single point
representing the system in its $2N$ dimensional phase space onto a
single $(x,v)$ plane, the system is represented by a cloud of $N$
points. If we let $N\rightarrow \infty$ while constraining the
total mass, $M=Nm$ , and energy $E$, the system is described by a
fluid in the $\mu (x,v)$ space with normalized distribution
function $f(x,v;t)$ satisfying the Vlasov (or Vlasov-Poisson)
equation

\begin{equation}\label{VlasEq}
\frac{\partial f}{\partial t}+v\frac{\partial f}{\partial
x}+A[f]\frac{\partial f}{\partial v}=0\ ,
\end{equation}
\begin{equation}
A[f]=2\pi MG\int dv^{\prime }\int dx^{\prime }f(x^{\prime
},v^{\prime },t)[\Theta (x^{\prime }-x)-\Theta (x-x^{\prime })]
\end{equation}

where $\Theta (x)$is the usual step function
\cite{vlasovia,2mass}. This system has been studied in the
literature extensively and is known to have an infinite number of
stationary solutions \cite{vlasovia,2mass,CHL}. In particular, the
maximum entropy solution is $f\sim\exp(-\beta
v^{2}/2)\cosh^{-2}(2\pi M^{2}Gx/2E)$.

\begin{figure}[tbp]
\centerline{\includegraphics[scale=0.7]{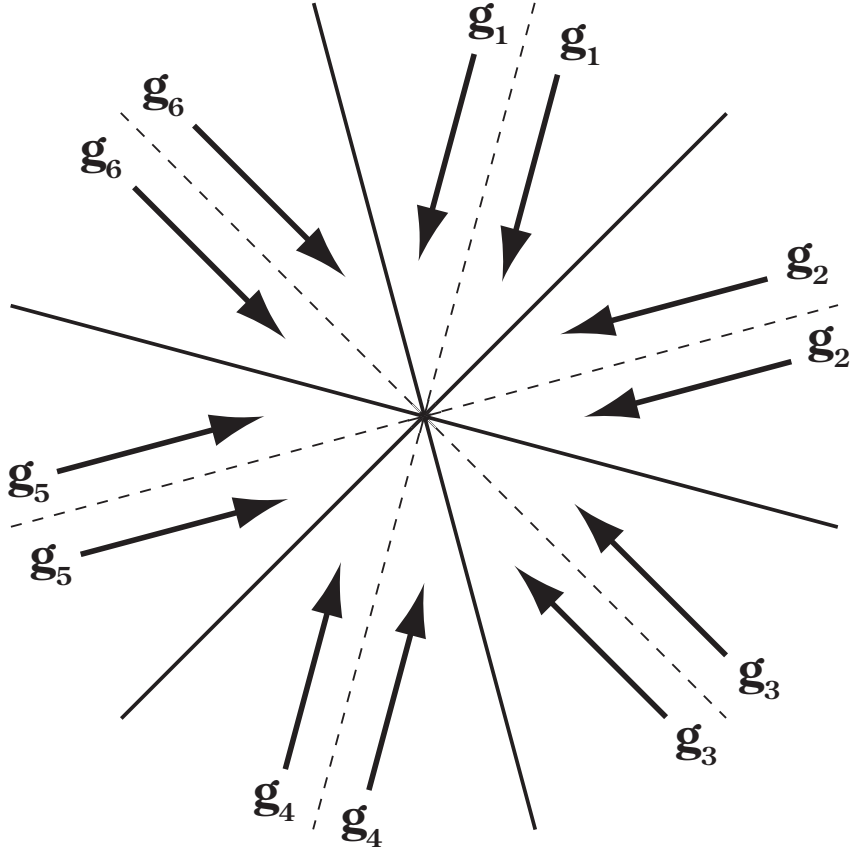}} \caption{Partition
of the plane into six segments. Arrows indicate direction of
acceleration. Note that within each wedge shaped segment the
acceleration is pointing inward and exactly parallel to the wedge
bisector.} \label{Pis6}
\end{figure}

To understand our approach, we first consider a simpler system of
a single particle moving in the plane in which we introduce the
usual polar coordinates, ($r,\theta )$. We imagine that the plane
is segmented into $P$ pie shaped ''wedges'' which share a common
vertex at the origin and subtend $2\pi/P$ radians. In the $k^{th}$
wedge the particle experiences a constant acceleration
$\mathbf{g}_{k}$, with $\left| \mathbf{g}_{k}\right|
=g$,($k=1,...,P)$ , directed along the wedge bisector, thus
pointing approximately towards the origin (common vertex, see Fig.
\ref{Pis6}). Then within each segment the particle executes a
parabolic trajectory until it reaches a wedge boundary, which it
crosses and begins a new parabolic segment with a rotated
acceleration. If, as in Fig. \ref{Pis6}, $P=6$ it has been shown
that this system is isomorphic to the system of three parallel
mass sheets \cite{wedge}. Note that the constraint on the center
of mass constrains the motion to the plane. We now imagine what
happens as $P$ becomes large. In each wedge the acceleration is
directed more nearly in the radial direction, i.e. towards the
origin. Thus the system is very close to another one where the
acceleration $\mathbf{g}=-g\mathbf{r}/r$ varies smoothly and is
defined everywhere except at the origin. This latter system is
exactly integrable: both energy and angular momentum are exactly
conserved. For large, but finite, $P$, most orbits in the former
system will lie close to their integrable counterparts for long
times. However, for any $P<\infty $, there will be time intervals
where $\frac{d\theta }{dt}$ changes sign for some of the orbits in
the segmented system and the orbits will start to diverge from
their integrable cousins.

Now let's shift our attention back to the OGS. Note that there are
$N!$ possible orderings of the particle labels. Associated with
each ordering is a pyramidal segment in the configuration space in
which the acceleration, $\mathbf{A}$, is a unique constant. To
explore how the geometry changes with increasing $N$, let's
compute $\cos (\phi ):=\mathbf{A}\cdot \mathbf{A}^{\ast }/A^{2}$
where $\mathbf{A}^{\ast }$ has the permuted particle ordering to
$\mathbf{A}$ obtained by interchanging the positions of a pair of
adjacent particles, and $A^{2}=\sum A_{j}^{2}$. With a little work
we obtain,

\begin{equation}\label{phi}
\frac{2(2\pi MG)^{2}}{N^{2}A^{2}}=\sin ^{2}(\phi /2)\ , \
A^{2}=\frac{1}{3}(2\pi MG)^{2}N(1-N^{-2}).
\end{equation}

Therefore, for $N\gg 1$, $\phi \cong 2\sqrt{6}/N^{3/2}$ becomes
vanishingly small, and we are faced with a similar situation to
the planar example discussed above, i.e. there are $N!$ cells
within which the acceleration is approximately in the radial
direction in the $N-1$ dimensional space obtained by projecting
out the center of mass direction $\mathbf{e}_{c}$. As a
consequence we can identify the OGS with a different system
consisting of a single particle moving in $N$ dimensions with a
radial acceleration of constant magnitude $A$. As we shall quickly
see, the latter is an exactly integrable system. Let us signify
the coordinates and velocities of the smooth system by
($y_{j},u_{j})$ so its position and velocity are $\mathbf{y=}\sum
y_{j}\mathbf{e}_{j}$, $\mathbf{u=} \sum u_{j}\mathbf{e}_{j}$. Then
the new equations of motion are

\begin{equation}\label{EqMo}
\frac{dy_{j}}{dt}=u_{j}\ ,\ \frac{du_{j}}{dt}=-\frac{Ay_{j}}{R},\
R^{2}=\sum\limits_{1}^{N}y_{j}^{2}.
\end{equation}

This is a central force problem in $N$ dimensions with a potential
energy $\Phi =mAR$.\ Direct substitution in the above quickly
shows that the energy, $H:=\frac{m}{2}u^{2}+\Phi $, the
generalized angular momentum two form
$\mathbf{L=}m\mathbf{y}\wedge \mathbf{v}$, and $L^{2}$ are exact
dynamical invariants of the motion. Moreover, if at an initial
time, $\mathbf{e}_{c}\mathbf{\cdot y=e }_{c}\mathbf{\cdot u}=0$,
then by taking successive derivatives of \ $\mathbf{y}$ and
$\mathbf{u}$ with respect to $t$ it can be seen that these
quantities don't change their value, and the trajectories remain
constrained to an $N-1$ dimensional manifold. It is remarkable
that, in spite of the arbitrary number of dimensions, since
$\frac{d\mathbf{u}}{dt}\sim \mathbf{y}$ , the motion remains in
the two-plane defined by the initial vectors $\mathbf{y}_{0}$ and
$\mathbf{u}_{0}$ ! Therefore the motion is simple to describe: If
we impose polar coordinates ($R,\theta $) in the plane of the
motion we obtain

\begin{equation}\label{PolEqMo}
\frac{d^{2}R}{dt^{2}}=-A+\frac{L^{2}}{m^{2}R^{3}}\quad ,\quad
\frac{d\theta }{dt}=\frac{L}{mR^{2}}
\end{equation}

Alternatively, a first order equation for $R(t)$,

\begin{equation}\label{Rad}
\frac{dR}{dt}=\pm \sqrt{\frac{2}{m}\left[
H-mAR-\frac{L^{2}}{2mR^{2}}\right] },
\end{equation}

can be obtained directly from the energy equation and shows that
turning points of the orbit occur where the argument of the
radical vanishes. Consequently $R(t)$ is a periodic function of
the time with period, say, $\tau _{R}$. However, we must be
careful to note that $\theta (t)$ does not advance by $2\pi $
radians in the time $\tau _{R}$ , so the orbits are not closed in
space. By inverting the equation (solving for $t=t(R)$) it can be
shown that the solution is a linear combination of elliptic
integrals.

To obtain the complete set of coordinates and velocities, we use
the initial position and velocity,
$\mathbf{y}_{0}=\mathbf{y}(t=0),$ $\
\mathbf{u}_{0}=\mathbf{u}(t=0)$, to define orthogonal unit vectors
in the plane, $\mathbf{b}_{1}$ , $\mathbf{b}_{2}:$

\begin{equation}\label{bbas}
\mathbf{b}_{1}=\frac{\mathbf{y}_{0}}{R_{0}}\ ,\
\mathbf{b}_{2}=\frac{(\mathbf{u}_{0}-(\mathbf{u}_{0}\cdot
\mathbf{e}_{y})\mathbf{e}_{y})}{(u_{0}^{2}-(\mathbf{u}_{0}\cdot
\mathbf{e}_{y})^{2})}.
\end{equation}

Then we can impose Cartesian coordinates ($z_{1},z_{2}$) such that

\begin{equation}\label{CarCo}
\mathbf{y}(t)=z_{1}\mathbf{b}_{1}+z_{2}\mathbf{b}_{2}\ ,\
\mathbf{u}(t)=\frac{dz_{1}}{dt}\mathbf{b}_{1}+\frac{dz_{2}}{dt}\mathbf{b}_{2}
\end{equation}

where ($z_{1,}z_{2})$ satisfy

\begin{equation}\label{CarCoMo}
\frac{d^{2}z_{1}}{dt^{2}}=-A\frac{z_{1}}{\sqrt{z_{1}^{2}+z_{2}^{2}}}\
,\
\frac{d^{2}z_{2}}{dt^{2}}=-A\frac{z_{2}}{\sqrt{z_{1}^{2}+z_{2}^{2}}}.
\end{equation}

The complete set of coordinates and velocities are obtained from
$y_{j}(t)=\mathbf{e}_{j}\cdot \mathbf{y}(t)$ ,
$u_{j}(t)=\mathbf{e}_{j}\cdot \mathbf{u}(t)$. Note that the
Cartesian pair ($z_{1,}z_{2}$) can be obtained either by solving
this pair of coupled second order equations, or by simply noting
that $z_{1}=R\cos \theta $, $z_{2}=R\sin \theta $.

Several continuum limits for the exactly integrable gravitational
analogue (EIS) can be constructed by focusing on $R^{2}$=$\sum
y_{j}^{2}$ . As for the OGS, project the phase space onto a single
two dimensional plane ($y,u$) to obtain a cloud of $\ N$ points,
and take the limit $N\longrightarrow \infty $ while constraining
the total mass, $M=Nm$ , and energy $H$. Let $f_{I}(y,u;t)$ be the
normalized density pdf in the $(y,u)$ plane. Then in the continuum
limit we can immediately identify $R^{2}=\sum y_{j}^{2}$ with
$N\int dy\int duy^{2}f_{I}:=N\left\langle y^{2}\right\rangle $
which, in general, depends on the time.We will see that in the
large $N$ \ limit the factors of $N$ will scale out of the
equations. Recalling that $A\sim \sqrt{N}$ , the equations of
motion of a point in the $(y,u)$ plane in the continuum limit are

\begin{equation}\label{EISMo}
\frac{dy}{dt}=u\ ,\ \frac{du}{dt}=-\frac{Ay}{\sqrt{N\left\langle
y^{2}\right\rangle }}=-\frac{(2\pi MG)y}{\sqrt{3\left\langle
y^{2}\right\rangle }},
\end{equation}

which results in the following Vlasov equation for $f_{I}(y,u,t)$:

\begin{equation}\label{EISVlas}
\frac{\partial f_{I}}{\partial t}+u\frac{\partial f_{I}}{\partial
y}-\frac{(2\pi MG)y}{\sqrt{3\left\langle y^{2}\right\rangle
}}\frac{\partial f_{I}}{\partial u}=0.
\end{equation}

It's important to keep in mind that, because of the dependence of
$\left\langle y^{2}\right\rangle$ on $f_{I}(y,u;t)$ (it is a
functional), this is not simply the equation of a distribution of
independent harmonic oscillators.

We have seen that explicit $N$ dependence disappears from the equations of
motion in the fluid limit. In the continuum limit it's convenient, and in
fact necessary, to introduce scaled quantities $r,h,a,l^{2},$

\begin{equation}\label{ScaVar}
r:=\frac{R}{\sqrt{N}}\ ,\ h:=\frac{H}{mN}\ ,\
a:=\frac{A}{\sqrt{N}}\ ,\ l^{2}:=\frac{L^{2}}{m^{2}N^{2}}.
\end{equation}

Then their limiting values are

\begin{equation}\label{ScaVar}
r^{2}=\left\langle y^{2}\right\rangle \ ,\
h=\frac{1}{2}\left\langle u^{2}\right\rangle +ar\ ,\ a=\frac{2\pi
MG}{\sqrt{3}}\ ,\ l^{2}=\left\langle y^{2}\right\rangle
\left\langle u^{2}\right\rangle -\left\langle yu\right\rangle
^{2}.
\end{equation}

A remarkable feature of the fluid limit is that the radial
equation is still preserved, i.e. substituting the scaled
variables gives the identical differential equation in terms of
the scaled variables, so that $r(t)$ is still a periodic function
of time. With a little work, we find from the Vlasov equation for
$f_{I}$ that both $h$ and $l^{2}$ are integral invariants of the
fluid motion defined by Eq. (\ref{EISMo}). In common with the OGS,
there are an infinite number of stationary solutions of
Eq.(\ref{EISVlas}). We can construct a family of such solutions by
defining the combined energy-like function (and functional) $g:$

\begin{equation}\label{EISFun}
g(y,u,[f_{I}]):=\frac{1}{2}u^{2}+\frac{1}{2}\frac{ay^{2}}{\sqrt{\left\langle
y^{2}\right\rangle }}.
\end{equation}

Note that $\frac{\partial g}{\partial u}$\ is the local velocity
and -$\frac{\partial g}{\partial y}$ is the local acceleration.
Therefore any integrable functional of $g$ satisfies the Vlasov
equation and yields a stationary solution $f_{I}[g]$. Note also
that $\left\langle g\right\rangle \neq \left\langle h\right\rangle
$. It is also possible to use the invariance of $h$ and $l^{2}$ to
construct the family of maximum entropy solutions.

The standard derivation of the Vlasov equation starts with the
BBGKY hierarchy for a system of pairwise interacting particles and
proves that, in the Vlasov limit, the higher order distribution
functions factor into products of the singlet distribution $f$
\cite{braunhepp}, so that pairs of particles become statistically
independent. In fact, we implicitly made this assumption in
determining the expression for $l^{2}$. However, in contrast with
the OGS, the EIS\ does not have pairwise additive interactions so
it may be possible to construct additional continuum
representations which include correlations. This possibility is
suggested by the form of the generalized angular momentum,
Eq.(\ref{AngMom}), which depends on the coordinates and velocities
of pairs of particles. Note that, in the continuum limit, both
$l_{p}:=(yu^{\prime }-y^{\prime }u)$ and its square are invariants
of the motion. Let $f_{I}^{(2)}(x,u;x^{\prime },u^{\prime };t)$ be
a normalized distribution in the four dimensional pair space and
assume that it satisfies the extended Vlasov equation

\begin{equation}\label{EISPairVlas}
\frac{\partial f_{I}^{(2)}}{\partial t}+u\frac{\partial
f_{I}^{(2)}}{\partial y}+u^{\prime }\frac{\partial
f_{I}^{(2)}}{\partial y^{\prime }}- \frac{(2\pi
MG)y}{\sqrt{3\left\langle y^{2}\right\rangle }}\frac{\partial
f_{I}^{(2)}}{\partial u}-\frac{(2\pi MG)y^{\prime
}}{\sqrt{3\left\langle y^{\prime 2}\right\rangle }}\frac{\partial
f_{I}^{(2)}}{\partial u^{\prime }}=0.
\end{equation}

Then, as in the standard formulation, one class of solutions can
always be constructed by the product $f_{I}(y,u;t)f_{I}(y^{\prime
},u^{\prime };t)$. However, there are now other possibilities. For
example, the integrable functionals of \ $l_{p}$ (e.g.
$f_{I}^{(2)}\sim \exp (-cl_{p}^{2})$) will generate a family of
stationary solutions in which pairs of points are obviously
correlated. It will be interesting to look for connections with
earlier studies of correlation in the OGS \cite{Monaghan}.

Of course, due to their dynamical instability, nearly all orbits
generated by the OGS will diverge from their EIS counterparts
sooner or later. To get an idea of how well the orbits generated
by the EIS follow the those of the OGS, we carried out numerical
simulations of the OGS and integrated the radial equation
numerically for the EIS. For the initial conditions we selected
random initial positions and velocities that are uniformly
distributed in a rectangle in $(x,v)$. We chose units of mass and
time where $M:=mN=1$ and $2\pi G=1.$We made small translations to
guarantee that the total momentum and center of mass are null, and
scaled the positions and velocities so that the total energy is
exactly 0.75 and the virial ratio (twice the kinetic energy
divided by the potential energy) is exactly $1.0$. In each case we
computed the initial values of $R^{2}$ and $L^{2}$ which are all
that is required to solve the radial equation, Eq.(\ref{Rad}).

\begin{figure}[tbp]
\centerline{\includegraphics[scale=0.8]{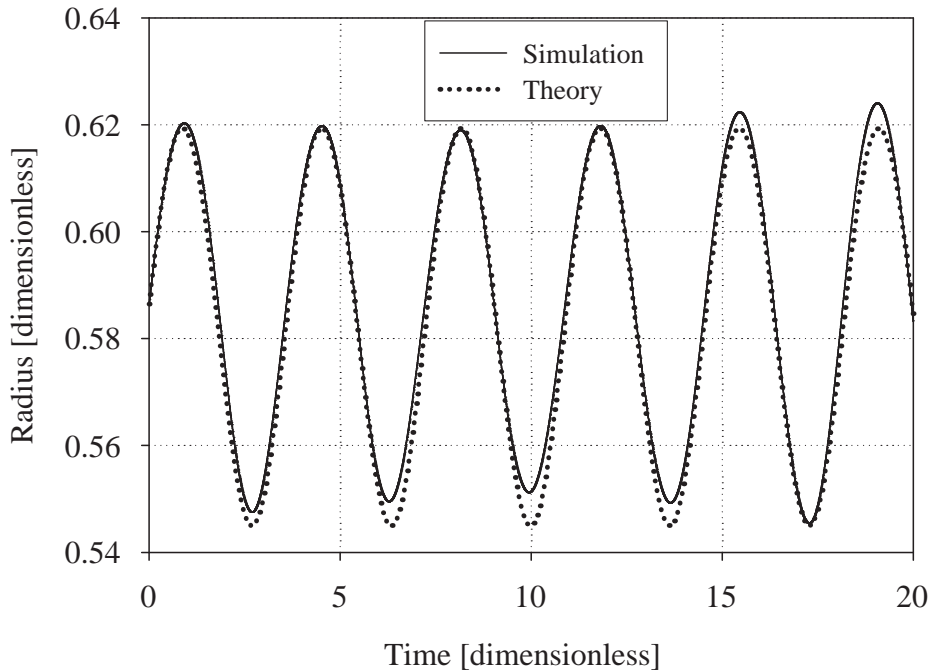}}
\caption{Comparison of the radial coordinate $R(t)$ computed for
the OGS and EIS with the same initial condition. Note the close
agreement for this initial condition.} \label{Reg_vs_Theory}
\end{figure}

\begin{figure}[tbp]
\centerline{\includegraphics[scale=0.8]{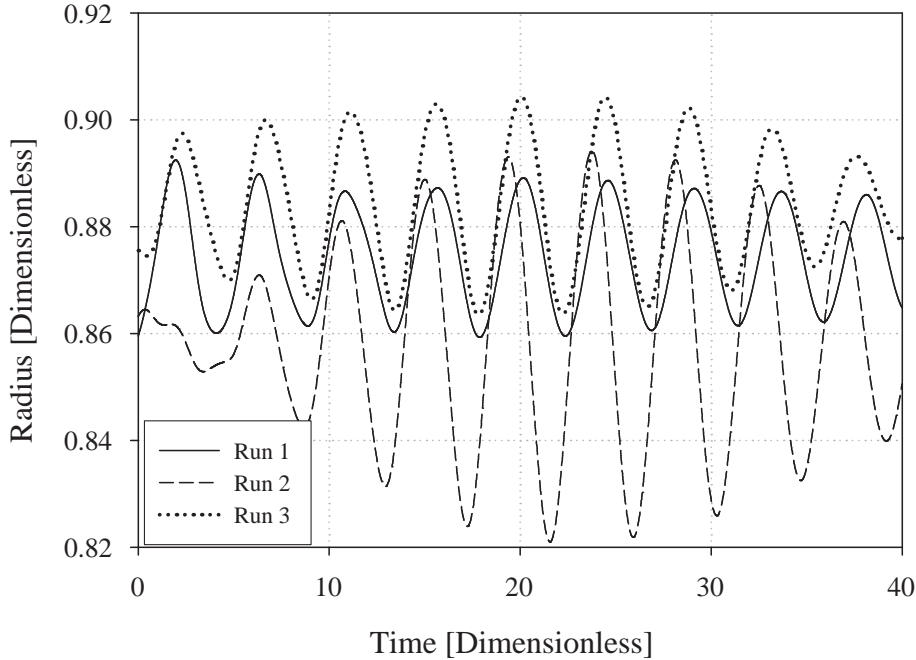}}
\caption{Evolution of the radial coordinate for the OGS from three
different initial conditions selected using the identical
algorithm.} \label{Random_Waterbag}
\end{figure}

Below we present graphs comparing the theory, from the EIS, and
simulations. Although the results are not exhaustive, there seems
to be a well established trend. In cases where the simulations
exhibit a very regular variation of the radius, there is a
remarkable agreement between theory (as given by the EIS) and
simulation. This can be seen in Figure \ref{Reg_vs_Theory}. For
other, equally likely, initial conditions the simulated $R(t)$ is
not regular and, therefore, is not well represented by the theory
(see Figure \ref{Random_Waterbag}). However, even in these cases,
we find that the maxima and minima of the simulations line up
fairly closely with those of the theory, i.e. they are in phase.
This seems to be a general trend whether we are modelling the
initial behavior or letting the system anneal for a while before
comparisons are made.

Starting from the $N$ particle one-dimensional self-gravitating
system (OGS), we have shown how to construct an exactly integrable
Hamiltonian system of arbitrary dimension (EIS) in which the
motion is always confined to a plane in the configuration space.
The system consists of a single point which experiences an
acceleration of constant magnitude that is always directed towards
a fixed point (say the origin) of $N$ dimensional Euclidean space.
The exact solution of the equations of motion can be expressed as
a linear combination of elliptic integrals. In general, the motion
fills an annulus in the plane between two turning points, but
degenerate cases are possible where the motion is exactly circular
(maximum possible angular momentum), resulting in stationary
motion, or linear (null angular momentum), resulting in the
periodic collapse and expansion of the system. By imposing the
Vlasov limit, we have shown how to construct two types of
continuum models for the EIS, one of which is capable of
representing correlations between particles. Simulations of the
OGS have been compared with solutions of the EIS with identical
initial conditions. For cases where the dynamical simulations are
regular, the agreement between the two models is outstanding. For
other less regular behavior, extrema in the simulations seem to
occur at similar times as the periodic extrema of the integrable
model, so the solutions are in phase. From the preliminary results
obtained so far, it appears that the existence of substructure in
the gravitational system is responsible for the lack of
regularity. Further work concerning the influence of population,
the graininess of distributions, and the theory and possible
applications of the continuum representations of the EIS is
ongoing. In addition we are investigating the extension of the
theory to higher dimensional systems.

\acknowledgements The authors benefitted from conversations with
John Hopkins, George Gilbert, Tony Burgess, and Igor Prokhorenkov,
as well as the support of the Research Foundation and Department
of Information Services of Texas Christian University.

\bibliographystyle{apsrev}
\bibliography{gravbib}

\end{document}